\documentstyle[amsfonts,12pt]{article}

\def\half{\textstyle{\frac{1}{2}}}

\def\H{{\cal H}}

\def\l{\lambda}

\def\ep{\epsilon}

\def\ra{\rightarrow}
\def\tint{{\textstyle\int}}

\def\H{{\cal H}}

\def\s{\hskip.08em}
\def\d{\partial}
\def\o{\overline}

\def\b{\begin{eqnarray*}}     %takes no eqn numbers
\def\e{\end{eqnarray*}}       %takes no eqn numbers
\def\bn{\begin{eqnarray}}     %takes eqn numbers 
\def\en{\end{eqnarray}}       %takes eqn numbers
\def\<{\langle}
\def\>{\rangle}

\def\no{\nonumber}

\def\{{\lbrace}
\def\}{\rbrace}
\begin{document}
%\footnote{Electronic mail: klauder@phys.ufl.edu}
\title{The Feynman Path Integral: \\
An Historical Slice}
\author{John R. Klauder
\footnote{Electronic mail: klauder@phys.ufl.edu}\\
Departments of Physics and Mathematics\\
University of Florida\\
Gainesville, FL  32611}
%Email: klauder@phys.ufl.edu}
\date{}     %   Use   %\date{} to see the dates
\maketitle
\begin{abstract}
Efforts to give an improved mathematical meaning to Feynman's path integral 
formulation of quantum mechanics
started soon after its introduction and continue to this day. In the present 
paper, one common thread of
development is followed over many years, with contributions made by various 
authors. The present version
of this line of development involves a continuous-time regularization for a 
general phase space path integral
and provides, in the author's opinion at least, perhaps the optimal 
formulation of the path integral. 
\end{abstract}
%\vfill\eject

\section*{The Feynman Path Integral, 1948}
Much has already been written about Feynman path integrals, and, no doubt, 
much more will be written in the future.
A comprehensive survey after more than fifty years since their introduction 
would be a major undertaking, and
this paper is not such a survey. Rather, it is an attempt to follow one 
relatively narrow development regarding a 
special form of regularization used in the definition of path integrals. 
Since we deal with several different approaches, this paper does not go 
too deeply into any one of them; it is intended more as a conceptual 
overview rather than a detailed exposition. In order to set the stage, 
let us start our discussion 
with a review of the traditional approach to path integral construction.

We begin with the Schr\"odinger equation
  \bn  i\hbar\s \frac{\d\psi(x,t)}{\d t}=-\frac{\hbar^2}{2\s m}
\s\frac{\d^2\psi(x,t)}{\d x^2}+V(x)\s\psi(x,t)  \en
appropriate to a particle of mass $m$ moving in a potential $V(x)$, 
$x\in{\mathbb R}$. A solution to this equation can be written as
an integral,
  \bn  \psi(x'',t'')=\int K(x'',t'';x',t')\s\psi(x',t')\,dx'\;,  \en
which  represents the wave function $\psi(x'',t'')$ at time $t''$ as a 
linear superposition over the wave function $\psi(x',t')$ at the initial 
time
$t'$, $t'<t''$. The integral kernel $K(x'',t'';x',t')$ is known as the 
{\it propagator}, and according to Feynman \cite{fey}
it may be given by
  \bn  K(x'',t'';x',t')={\cal N}\int e^{(i/\hbar)\tint[(m/2)\s 
{\dot x}^2(t)-V(x(t))]\,dt}\;{\cal D}x\;,  \en
which is a formal expression symbolizing an integral over a suitable 
set of paths. This integral is supposed to run over
all continuous paths $x(t)$, $t'\le t\le t''$, where $x(t'')=x''$ and 
$x(t')=x'$ are fixed end points for all paths. Note that the integrand 
involves the
classical Lagrangian for the system. 

Unfortunately, although highly suggestive, the preceding expression for 
the path integral is {\it undefined} as it
stands: For example, the normalization constant ${\cal N}$ diverges, and 
the putative translation invariant measure
${\cal D}x$ does not exist. To overcome these basic problems, Feynman 
adopted a {\it lattice regularization} as a
procedure to yield well-defined integrals which was then followed by a 
limit as the lattice spacing goes to zero called the continuum limit. 
With $\epsilon>0$
denoting the lattice spacing, the details regarding the lattice 
regularization procedure are given by
  \bn &&K(x'',t'';x',t')=\lim_{\ep\ra0}\s(m/2\pi i\hbar\ep)^{(N+1)/2}
\s\int\cdots\int \no\\
&&\hskip1cm\times\exp\{(i/\hbar)\Sigma_{l=0}^N\s [(m/2\ep)(x_{l+1}-x_l)^2
 -\ep\s
V(x_l)\s]\}\;\Pi_{l=1}^N\,dx_l \;,  \en
where $x_{N+1}=x''$, $x_0=x'$, and $\ep\equiv(t''-t')/(N+1)$, $N\in\{1,2,3,
\dots\}$.
In this version, at least, we have an expression that has a reasonable 
chance of being well defined, provided, of course, 
that one interprets the conditionally convergent integrals involved in an 
appropriate manner. One common and fully
acceptable interpretation adds a convergence factor to the exponent of 
the preceding integral in the form  
  \bn -(\ep^2/2\hbar) \Sigma_{l=1}^N\s x^2_l\;,  \en
which is a term that formally makes no contribution to the final result 
in the continuum limit save for ensuring that the integrals 
involved are now rendered absolutely convergent. 

Accepting the fact that the integrals all converge ensures that a 
meaningful function of $\ep$ emerges, but, by itself,
that fact does not ensure convergence as $\ep\ra0$ and, even if 
convergence holds, it furthermore does not guarantee
that the result is correct! To ensure convergence to the correct 
result requires that some technical condition(s) must be imposed
on the potential $V(x)$. In this regard, we only observe that a correct 
result emerges whenever the potential has
a lower bound, i.e., whenever $V(x)\ge c$, for some $c$, $-\infty<c<\infty$, 
for all $x$.

We recall that for the free particle of mass $m$ the potential 
$V(x)=0$ for all $x$. In that case, Eq.~(4) reads 
  \bn &&\hskip-.9cm K(x'',t'';x',t')=\lim_{\ep\ra0}\s(m/2\pi 
i\hbar\ep)^{(N+1)/2}\s\int\cdots\int \no\\
&&\hskip2.5cm\times\exp\{(i/\hbar)\Sigma_{l=0}^N\s 
[(m/2\ep)(x_{l+1}-x_l)^2
\}\;\Pi_{l=1}^N\,dx_l \no\\
&&\hskip1.8cm =\sqrt{\frac{m}{2\pi i\hbar(t''-t')}}\,
\exp{\bigg(\frac{im(x''-x')^2}{2\hbar(t''-t')}\bigg)}\;,  \en 
which is the form of the quantum mechanical propagator for the free particle.

\subsection*{Comments}
The procedure sketched above --- whereby the action functional 
expressed as an integral over a continuous-time parameter
is replaced by a natural Riemann sum approximation, which eventually 
is followed by a continuum limit ($\ep\ra0$) as the
final step --- provides a satisfactory procedure for a suitable and 
large class of potentials to define the
propagator $K(x'',t'';x',t')$. However, it is important to stress 
that a lattice regularization followed by a continuum limit is by 
no means the only way 
to give a satisfactory definition of a path integral, nor, in the 
author's opinion does it even represent the most satisfactory 
definition, although admittedly this latter issue is in part subjective. 
What follows in this paper is a narrow
review --- an historical slice --- of the development of various efforts 
to find a suitable {\it continuous-time 
regularization} procedure along with a subsequent limit to remove that 
regularization that ultimately should yield the correct propagator. 
While some of the work to be
described did not follow directly from work that preceded it, all of 
this work does, with appropriate hindsight,
seem to have a fairly natural progression that makes an interesting 
story.

\section*{Feynman-Kac Formula, 1951}
Through his own research, Mark Kac was fully aware of Wiener's theory of 
Brownian motion and the associated diffusion equation
that describes the corresponding distribution function. Therefore, it 
is not surprising that he was well prepared to
give a path integral expression in the sense of Feynman for an equation 
similar to the time-dependent Schr\"odinger equation save for a
rotation of the time variable by $-\pi/2$ in the complex plane, namely, 
by the change $t\ra-it$. In particular, Kac \cite{kac}
considered the equation
  \bn  \frac{\d\rho(x,t)}{\d t}=\frac{\nu}{2}\s\frac{\d^2\rho(x,t)}
{\d x^2}-V(x)\s\rho(x,t) \;. \label{t7}\en
This equation is analogous to Schr\"odinger's equation but of course 
differs from it in certain details. Besides
certain constants which are different, and the change $t\ra-it$, the 
nature of the dependent variable function
$\rho(x,t)$ is quite different from the normal quantum mechanical wave 
function. For one thing, if the function $\rho$ is initially real it 
will remain real as time
proceeds. Less obvious is the fact that if $\rho(x,t)\ge0$ for all $x$ 
at some time $t$, then the function will continue 
to be nonnegative for all time $t$. Thus we can interpret $\rho(x,t)$ 
more like a probability density; in fact in the special case that $V(x)=0$,
then $\rho(x,t)$ is the probability density for a Brownian particle which 
underlies the Wiener measure. In this regard, $\nu$ is called the 
diffusion constant. 

The fundamental solution of Eq.~(\ref{t7}) with $V(x)=0$ is readily given as
  \bn W(x,T;y,0)=\frac{1}{ \sqrt{2\pi\nu T}}\, 
\exp{\bigg( -\frac{(x-y)^2}{2\nu T}\bigg)}\;,  \en
which describes the solution to the diffusion equation subject to the 
initial condition 
  \bn \lim_{T\ra0^+}\s W(x,T;y,0)=\delta(x-y)\;. \en
Moreover, it follows that the solution of the diffusion equation for a 
general initial condition is given by
  \bn \rho(x'',t'')=\int W(x'',t'';x',t')\s \rho(x',t')\,dx'  \;. \en
Iteration of this equation $N$ times, with $\ep=(t''-t')/(N+1)$, leads 
to the equation
  \bn  \rho(x'',t'')=N'\int\cdots\int e^{-(1/2\nu\ep)
\Sigma_{l=0}^N(x_{l+1}-x_l)^2}\;\Pi_{l=1}^N dx_l\;\rho(x',t')\,dx'\;,  \en
where $x_{N+1}\equiv x''$ and $x_0\equiv x'$. This equation features 
the imaginary time propagator for a free particle of unit mass as
given by
  \bn  W(x'',t'';x',t')=N'\int\cdots\int e^{-(1/2\nu\ep)
\Sigma_{l=0}^N(x_{l+1}-x_l)^2}\;\Pi_{l=1}^N dx_l\;.  \en
Since this equation holds for all $N>0$, we may assume that it also 
holds in the limit $\ep\ra0$, i.e., $N\ra\infty$, which we can
write formally as
   \bn W(x'',t'';x',t')={\cal N}\int e^{-(1/2\nu)\tint {\dot x}^2\,dt}
\;{\cal D}x\;, \en
where ${\cal N}$ denotes a formal normalization factor. (Symbols such as 
${\cal N}$ may stand for different factors in different expressions.)

The similarity of this expression with the Feynman path integral 
[for $V(x)=0$] is clear, but there is a profound difference between these
equations. In the former (Feynman) case the underlying measure is only 
{\it finitely additive}, while in the latter (Wiener) case the continuum 
limit actually defines a genuine measure, i.e., a {\it countably 
additive measure} on paths, which is a version of the justly famous
Wiener measure. In particular, 
  \bn W(x'',t'';x',t')=\int d\mu_W^\nu(x)\;,  \en
where $\mu_W^\nu$ denotes a measure on continuous paths $x(t)$, 
$t'\le t\le t''$, for which $x(t'')\equiv x''$ and  $x(t')\equiv x'$.
Such a measure is said to be a {\it pinned} Wiener measure, since it 
specifies its path values at two time points, i.e., at $t=t'$
and at $t=t''>t'$. (The traditional Wiener measure, which we shall not 
really deal with in this paper, specifies its values only at the initial 
time,
and it corresponds to one additional integration over the final value 
$x''$ at the final time $t''$.)

We note without proof that Brownian motion paths have the property that 
with probability one they are concentrated on continuous paths. However, 
it is also true that the time derivative of a Brownian path is almost 
nowhere defined, which means that, with probability one, 
${\dot x}(t)=\pm\infty$ for all $t$.

When the potential $V(x)\neq0$ the propagator associated with (7) is 
formally given by
  \bn W(x'',t'';x',t')={\cal N}\int e^{-(1/2\nu)
\tint {\dot x}^2\s dt-\tint V(x)\s dt}\;{\cal D}x\;, \en
an expression which is well defined if $V(x)\ge c$, $-\infty<c<\infty$.
A mathematically improved expression for (15) makes use of the Wiener 
measure and is given by
  \bn W(x'',t'';x',t')=\int e^{-\tint V(x(t))\s dt}\;d\mu^\nu_W(x)\;. \en
This is an elegant relation in that it represents a solution to the 
differential equation (7) in the form of an integral over Brownian 
motion paths suitably weighted by the potential $V$. Incidentally, 
since the propagator is evidently a strictly positive function, it follows 
that the solution of the differential equation (7) is nonnegative for all 
time $t$ provided it is nonnegative for any particular time value.

\section*{Gel'fand and Yaglom, 1956}
In an important review article \cite{gel}, these two authors introduced 
the concept of a continuous-time regularization of Feynman path integrals. 
Although their procedure was later shown to be incorrect, their work may 
be said to have initiated an interesting line of development.

The idea of Gel'fand and Yaglom was relatively straightforward. Formally 
stated, their proposal was to define the propagator as 
   \bn  \lim_{\nu\ra\infty}{\cal N}_\nu\int \exp\{(i/\hbar)
\tint[(m/2){\dot x}^2-V(x)]\s dt\}\exp\{-(1/2\nu)\tint{\dot x}^2\s dt\}
\;{\cal D}x\;. \en
Formally, therefore, their proposal consisted of introducing an auxiliary 
factor
into the integrand that is identical to the factor which leads to Wiener 
measure. The purpose of such a factor is to introduce a regularization 
into the original formal expression. Finally, the limit $\nu\ra\infty$ 
is taken as a last step which amounts to formally removing the auxiliary 
factor and leaving behind the original integrand of the Feynman path integral.

There is one glaring difficulty with this proposal. The auxiliary factor 
which leads to Brownian motion paths with a diffusion constant 
$\nu<\infty$ has the property of ensuring that the paths are continuous, 
but they are nowhere differentiable. For such paths, the contribution of 
the potential is well defined, but the contribution of the kinetic energy 
is divergent for all paths. Gel'fand and Yaglom hoped to get around this 
difficulty by combining the kinetic energy with the corresponding factor 
in the auxiliary term giving rise to a Wiener measure with a 
{\it complex} diffusion coefficient $\sigma$, where
  \bn \frac{1}{\sigma}=\frac{1}{\nu}-\frac{i\s m}{\hbar}\;. \en
Thus the strategy was to take the limit of a sequence of presumably well 
defined integrals over Wiener measure with a complex diffusion constant 
$\sigma$ as the real part of $\sigma$ vanished. Notice that this kind of 
regularization does not involve the introduction of a temporal lattice 
followed by a continuum limit but rather maintains a continuous time 
parameter throughout. It is for this reason that we refer to this kind 
of procedure as a {\it continuous-time regularization scheme}. 

\section*{Cameron, 1960}
Shortly after the appearance of the Gel'fand and Yaglom paper, Cameron 
\cite{cam} showed that the scheme was fatally flawed. In particular, 
Cameron showed that a Wiener measure defined with a complex diffusion 
constant $\sigma$
leads to a countably additive measure only when $\sigma\equiv{\rm Re}
\s\sigma>0$, hence ${\rm Im}\s\sigma\equiv0$. It is
instructive to sketch the argument that leads to this conclusion. Consider 
the
$N$-fold integral
  \bn  &&\bigg(\frac{\lambda}{2\pi\ep}\bigg)^{(N+1)/2}\int\cdots\int 
\exp[-(\lambda/2\ep)\s\Sigma_{l=0}^N\s(x_{l+1}-x_l)^2]\;\Pi_{l=1}^N\,dx_l
\no\\
  &&\hskip1cm=\bigg(\frac{\lambda}{2\pi N\ep}\bigg)^{1/2}\s 
\exp[-(\lambda/2N\ep)\s(x_{N+1}-x_0)^2]\;,  \en
which represents one of the primary formulas involved in discussing Wiener
measures. This formula holds for any complex $\lambda$ so long as 
${\rm Re}\s\lambda>0$.
If such a formula exists in the limit that $\ep\ra0$ and $N\ra\infty$ 
such that
$N\ep$ remains constant, then the desired path measure will also exist. 
Although that condition would appear to be evident, it is not true for 
general $\lambda$.

Integrals exist when they converge absolutely. Absolute convergence 
certainly
holds for (19) for all $N<\infty$, but what about when $N\ra\infty$? To 
test this issue, we consider the absolute integral associated with (19) 
given by
 \bn  &&\bigg(\frac{|\lambda|}{2\pi\ep}\bigg)^{(N+1)/2}\int\cdots\int 
\exp[-({\rm Re}\s\lambda/2\ep)\s\Sigma_{l=0}^N\s(x_{l+1}-x_l)^2]\;
\Pi_{l=1}^N\,dx_l\no\\
  &&\hskip1cm=\bigg(\frac{|\lambda|}{{\rm Re}\s\lambda}\bigg)^{N/2}\bigg(
\frac{|\lambda|}{2\pi N\ep}\bigg)^{1/2}\s \exp[-({\rm Re}\s\lambda/2N\ep)
\s(x_{N+1}-x_0)^2]\;.  \en
Evidently, the final result exists as $N\ra\infty$ only provided $|\lambda|
\equiv{\rm Re}\s\lambda>0$, i.e., provided ${\rm Im}\s\lambda\equiv0$, 
as claimed.
Consequently, the proposal of Gel'fand and Yaglom fails on the ground that 
whenever the diffusion constant for the Wiener measure is truly complex 
(${\rm Im}\s\sigma\neq0$), then the so-defined Wiener measure is only 
{\it finitely additive and not countably additive}. In this case, finite 
additivity means that finite answers for general integrals involving this 
measure arise by delicate cancellation of both positive and negative 
divergent contributions. A regularization scheme that results in a 
finitely additive measure on paths is not qualitatively different from 
the original formal Feynman path integral, and therefore it does not 
constitute an acceptable continuous time regularization.

\section*{It\^o, 1962}
Soon after Cameron pointed out the lack of a countably additive measure for 
the Gel'fand-Yaglom procedure, It\^o \cite{ito} proposed another version of 
a continuous-time regularization that resolved some of the troublesome 
issues. In essence, the proposal of It\^o takes the form given by 
  \bn \lim_{\nu\ra\infty}{\cal N}_\nu\int \exp\{(i/\hbar)\tint[\half 
m{\dot x}^2-V(x)]\s dt\}\exp\{-(1/2\nu)\tint[{\ddot x}^2+{\dot x}^2]
\s dt\}\;{\cal D}x\;.  \en
Note well the alternative form of the auxiliary factor introduced as a 
regulator. The additional term ${\ddot x}^2$, the square of the second 
derivative of $x$, acts to smooth out the paths sufficiently well so 
that in the case of (21) both $x(t)$ and ${\dot x}(t)$ are continuous 
functions, leaving ${\ddot x}(t)$ as the term which does not exist. 
However, since only $x$ and ${\dot x}$ appear in the rest of the integrand, 
the indicated path integral can be well defined; this is already a positive 
contribution all by itself! 

To proceed further with (21) we need to decide on what values to fix at 
the initial and final times. As was the case with the procedure proposed by 
Gel'fand and Yaglom, let us first suppose that a composition law of the 
expression in (21) holds {\it before} the limit $\nu\ra\infty$ is taken. 
This requires that one fix not only $x(t'')=x''$ and $x(t')=x'$ but, in 
addition, that we fix ${\dot x}(t'')={\dot x}''$ and ${\dot x}(t')=
{\dot x}'$. Viewed as a solution of the Schr\"odinger equation, however, 
we are at a loss to understand the meaning of the values of $x$ and 
${\dot x}$ to be held at any time. Thus, while this interpretation may 
yield a limiting function it is hard to see that such a result could be 
a solution to Schr\"odinger's equation. 

In point of fact, It\^o did not choose the previous set of data to be 
fixed  but he chose another set. It\^o interpreted (21) as an integral 
in which $x(t'')=x''$ and $x(t')=x'$ are the only values held fixed. He 
therefore did not require any composition law on the expression (21) at 
finite $\nu$, but only expected a composition law to hold for the 
expression that arises after $\nu\ra\infty$. 

With such an interpretation in mind we proceed to evaluate in a fairly 
heuristic manner the functional integral (21). For simplicity we shall 
discuss only the simple case where $V(x)=0$; moreover, we set $t'=0$, 
$t''=T$, and choose $x(0)=0$, $x(T)=x$. Thus, prior
 to taking the limit $\nu\ra\infty$, let us focus on the formal integral
  \bn {\cal N}_\nu \int\exp\{(i/\hbar)\tint\s {\dot x}\s g(t)\, dt\}\,
\exp\{\s-(1/2\nu)\tint[\s{\ddot x}^2+a^2{\dot x}^2\s]\,dt\}\,\delta(x(0))
\,{\cal D}x\;, \en
where $a^2\equiv 1-im\nu/\hbar$, $g(t)$ will be chosen below, and for the 
foregoing expression we assume there is no other pinning; the required 
second pinning at $t''=T$ will be introduced in a moment. Let us set 
 \bn x(t)\equiv \tint_0^t\,\xi(u)\,du\;, \en
for all $t$, so that the former integral can be replaced by
 \bn {\cal N}_\nu\int\exp\{(i/\hbar)\tint\s\xi\s g(t)\,dt\}\,\exp\{\s-(1/2\nu)
\tint[{\dot\xi}^2+a^2\xi^2\s]\,dt\}\,{\cal D}\xi\;. \en
In this form, the integral involves an Ornstein-Uhlenbeck measure 
\cite{orn}, and the answer is readily given by
  \bn \exp[\s-(\nu/4a\hbar^2)\tint g(t)\s g(u)\s e^{-a|t-u|}\;dt\,du]\;, \en
where we have chosen ${\cal N}_\nu$ so that the answer is 
unity if $g(t)=0$ for all $t$.
As a next step we let $g(t)=\l$ for $0\le t\le T$, and $g(t)=0$ otherwise. 
Thus (25) becomes
  \bn \exp[\s-(\nu\l^2/4a\hbar^2)\tint_0^T\tint_0^T\s e^{-a|t-u|}\;dt\,du]
\;. \en
Since 
  \bn &&\int e^{(i/\hbar)\l\tint_0^T \xi(t)\s dt}\, e^{-(i/\hbar)\l\s x}
\,d\l/(2\pi\hbar)\no\\
&&\hskip1cm =\delta(\tint_0^T\xi(t)\s dt-x)\no\\
&&\hskip1cm = \delta(x(T)-x)\;, \en
it follows that
 \bn &&{\cal N}_\nu\int\delta(x(T)-x)\,e^{-(1/2\nu)
\tint[{\dot\xi}^2+a^2\xi^2]\s dt}\,{\cal D}\xi\no\\
&&\hskip1cm =\int e^{-(i/\hbar)\l\s x}\,e^{-\nu\l^2 F/4a\hbar^2}
\,d\l/(2\pi\hbar)\no\\
&&\hskip1cm =\sqrt{\frac{a}{\nu F\pi}}\,
\exp{\bigg(-\frac{ax^2}{\nu F}\bigg)}\;, \en
where
  \bn &&F\equiv \tint_0^T\tint_0^T\,e^{-a|t-u|}\,dt\,ds\no\\
  &&\hskip.445cm =\frac{2T}{a}-\frac{2}{a^2}(1-e^{-aT})\;. \en
As $\nu$ becomes large, we may set $a^2\simeq -im\nu/\hbar$, in which case
$a/\nu F$ may be replaced by $-im/2T\hbar$. Hence, we are led to the final 
result
 \bn \sqrt{\frac{m}{2\pi iT\hbar}}\,
\exp{\bigg(\frac{imx^2}{2\hbar T}\bigg)}\;, \en
which is recognized as the quantum mechanical propagator 
$K(x,T;0,0)$ for a free particle of mass $m$ as given in (6). 

It\^o \cite{ito} made this story rigorous for constant, linear, and quadratic
potentials, as well as those potentials of the form 
  \bn V(x)=\tint e^{ixs}\,w(s)\,ds \en
provided that $\tint |w(s)|\s ds<\infty$. 

In summary, by introducing a regularization involving a higher 
derivative $({\ddot x}^2)$, It\^o was able to soften the paths 
sufficiently to allow the kinetic energy to be well defined. In so 
doing he was able to give a satisfactory continuous-time regularization 
for (21) for a certain class of potentials $V$.

\section*{Feynman, 1951}
It is necessary to retrace history at this point to recall the 
introduction of the {\it phase space path integral} by Feynman \cite{fey2}. 
In Appendix B to this article, Feynman introduced a formal
expression for the configuration or $q$-space propagator given by
  \bn K(q'',t'';q',t')={\cal M}\int \exp\{(i/\hbar)
\tint[\s p\s{\dot q}-H(p,q)\s]\s dt\}\;{\cal D}p\,{\cal D}q\;. \en
In this equation one is instructed to integrate over all paths $q(t)$, 
$t'\le t\le t''$, with $q(t'')\equiv q''$ and $q(t')\equiv q'$ held fixed, 
as well as
to integrate over all paths $p(t)$, $t'\le t\le t''$, without restriction.
As customary, this is a formal statement and in practice it needs to be 
given a precise meaning. The lattice prescription in which the continuous 
time parameter is replaced by a finite set of discrete points is the 
procedure that is
typically followed. For completeness, we illustrate a common lattice space 
version of the formal phase space path integral expression, as given by
  \bn &&K(q'',t'';q',t')=\lim_{\ep\ra0}\int\cdots\int\exp\{(i/\hbar)
\Sigma_{l=0}^N[\s\half \s p_{l+1/2}(q_{l+1}-q_l)\no\\
&&\hskip1cm -\ep\s H(p_{l+1/2},\half(q_{l+1}+q_l))\s]\s\}\;\Pi_{l=0}^N
\s dp_{l+1/2}/(2\pi\hbar)\,\Pi_{l=1}^N\s dq_l\;. \en
In this expression, all $p$ and $q$ variables are integrated over except 
for $q_{N+1}\equiv q''$ and $q_0\equiv q'$, and, just as before, 
$\ep=(t''-t')/(N+1)$. Since $q_l$ implies a sharp $q$ value at time 
$t'+l\ep$, we have chosen to name the conjugate variable $p_{l+1/2}$ 
to emphasize that a sharp $p$ value must occur at a different time, here 
at $t'+(l+1/2)\ep$,  since it is not possible to have sharp $p$ and $q$ 
values at the same time. Note that there is one more $p$ integration 
than $q$ integration
in this formulation. This discrepancy becomes clear when one imposes the 
composition law which requires that
  \bn K(q''',t''';q',t')=\int K(q''',t''';q'',t'')\s K(q'',t'';q',t')
\,dq''\;, \en
a relation which implies, just on dimensional grounds, that there must be 
one more $p$ integration than $q$ integration in the definition of each 
$K$ expression.

Observe that (32) exhibits an apparent covariance under canonical 
coordinate transformations, but this is quite illusory. In fact, the 
lattice form (33) is correct only when the canonical variables are 
described by Cartesian coordinates, and this kind of limitation on 
straightforward lattice-space regularizations is quite general. In a 
later section of this paper we shall return to this point and explain 
why this limitation is necessary.

It is also instructive to consider an alternative phase space path integral 
for the momentum or $p$-space propagator formally given by
  \bn K(p'',t'';p',t')={\cal M}\int \exp\{(i/\hbar)\tint[\s-\s q
\s{\dot p}-H(p,q)\s]\s dt\}\;{\cal D}p\,{\cal D}q\;,  \en
which is obtained from (32) by Fourier transformation on both end variables, 
$q''$ and $q'$.
In this case a lattice space definition can be given by
  \bn &&K(p'',t'';p',t')=\lim_{\ep\ra0}\int\cdots\int\exp\{(i/\hbar)
\Sigma_{l=0}^N[\s-\half \s q_{l+1/2}(p_{l+1}-p_l)\no\\
&&\hskip1cm -\ep\s H(\half(p_{l+1}+p_l),q_{l+1/2})\s]\s\}\;\Pi_{l=1}^N
\s dp_{l}\,\Pi_{l=0}^N\s dq_{l+1/2}/(2\pi\hbar)\;, \en
and we see in this expression that there is one more $q$ integration than 
$p$ integration. Again this makes sense when one considers the composition law
  \bn K(p''',t''';p',t')=\int K(p''',t''';p'',t'')\s K(p'',t'';p',t')
\,dp''\;.\en

On the other hand, and following similar reasoning, one would be hard 
pressed to give a satisfactory lattice space formulation of the putative 
formal phase space path integral given by
  \bn {\cal M}\int \exp\{(i/\hbar)
\tint[\s\half(p{\dot q}-q{\dot p})-H(p,q)\s]\s dt\}\;{\cal D}p
\,{\cal D}q\;.  \en
(In the light of the previous discussion, even the physical meaning of 
such an expression seems uncertain, at least to the present author. On 
the other hand, such an expression is easily understood in the coherent 
state representations that follow.)

It is widely appreciated that the phase space path integral is more 
generally applicable than the original, Lagrangian, version of the 
path integral. For instance, the original configuration space path 
integral is satisfactory for Lagrangians of the general form
  \bn L(x)=\half\s m{\dot x}^2+A(x)\s {\dot x}-V(x)\;, \en
but it is unsuitable, for example, for the case of a relativistic 
particle with the Lagrangian
  \bn L(x)=-m\sqrt{1-{\dot x}^2\s} \en
expressed in units where the speed of light is unity.
For such a system --- as well as many more general expressions --- the 
phase space form
of the path integral is to be preferred. In particular, for the relativistic
free particle, the phase space path integral
  \bn {\cal M}\int\exp\{(i/\hbar)\tint[\s p\s{\dot q}-\sqrt{p^2+m^2}\s]
\s dt\}\;{\cal D}p\,{\cal D}q\;,  \en
interpreted in the sense of (33), is readily evaluated and yields the 
correct propagator. 

Issues of proper coordinate choice also arise for the configuration space 
path integral as well, and expressions such as (3) and (4) implicitly 
refer to Cartesian coordinates.

\section*{Coherent State Representations, 1960}
As a prelude to the following section it is pedagogically useful to recall 
some basic properties of coherent states and the Hilbert space 
representations they generate \cite{klacoh}. We focus on coherent 
states generated by Heisenberg canonical operators $P$ and $Q$ which 
obey the basic commutation relation $[Q,P]=i\hbar\s I$, where $I$ denotes 
the unit operator. More specifically, we shall assume the Weyl form of the 
commutation relation holds for self-adjoint momentum and position 
operators in the form
  \bn e^{-iq P/\hbar}\s e^{ip\s Q/\hbar} = e^{-ip\s q/2\hbar}
\s e^{i(p\s Q-q P)/\hbar}\equiv U[p,q]\;, \en
where $p$ and $q$ are arbitrary real variables. In an abstract Hilbert 
space, let us introduce a normalized fiducial vector $|\eta\>$, which 
is otherwise arbitrary, and define vectors of the form
  \bn |p,q\>\equiv U[p,q]\s|\eta\>  \en
for all $(p,q)\in{\mathbb R}^2$. These states are the {\it canonical 
coherent states}, and they have a number of interesting properties. 
Since $U[p,q]$ is a unitary operator, it follows that each vector is 
normalized, i.e.,
  \bn \||p,q\>\|\equiv\<p,q|p,q\>^{1/2}=1\;. \en
Next, it should be noted that the coherent states are continuously 
parameterized; specifically, if $(p,q)\ra(p',q')$ in the sense that 
$|p-p'|^2+|q-q'|^2\ra0$, then it follows for any choice of $|\eta\>$, 
that $\||p,q\>-|p',q'\>\|\ra0$, or as one says, the vectors are strongly 
continuous in the parameters $p$ and $q$. Finally, we want to stress the 
important property that these vectors admit a rather conventional looking 
resolution of unity as a superposition over one-dimensional projection 
operators onto the coherent states, and specifically that
  \bn \int \s|p,q\>\<p,q|\,dp\s dq/(2\pi\hbar)=I  \en
holds, where, as before, $I$ is the unit operator, and this relation holds 
for any choice of $|\eta\>$. It is easy to show that the resolution of 
unity holds weakly in the sense that, for arbitrary vectors $|\psi\>$ 
and $|\phi\>$, 
  \bn  \int \<\phi|p,q\>\<p,q|\psi\>\,dp\s dq/(2\pi\hbar)=\<\phi|\psi\>
\;. \en
(It is less easy to show, but nevertheless true, that the resolution of 
unity also holds as a strong operator identity.) 

The resolution of unity formula permits us to introduce novel 
representations of Hilbert space rather different from those customarily 
used, particularly in quantum mechanics. Let us introduce the functional 
representatives
  \bn \psi(p,q)\equiv\<p,q|\psi\> \;, \en
which are defined for all $|\psi\>$ in the Hilbert space. As we have 
already done for the coherent states themselves, we suppress any 
dependence of the functions $\psi(p,q)$ on the state $|\eta\>$. Indeed, 
for any choice of $|\eta\>$, each such function is a bounded, continuous 
function of the variables $p$ and $q$. For each pair of such functions, 
such as $\psi(p,q)=\<p,q|\psi\>$ and $\phi(p,q)=\<p,q|\phi\>$, we assign 
the inner product given by
  \bn (\phi,\psi)\equiv\int \phi(p,q)^*\s\psi(p,q)\,dp\s dq/(2\pi\hbar)
=\<\phi|\psi\>\;. \en
The last part of this relation shows that the functional inner 
product equals the inner product in the abstract Hilbert space. The 
so-defined functional representation is complete simply because every 
element of the abstract Hilbert space is imaged as a function of the 
form $\psi(p,q)$. It should be noticed that the space of functions defined 
in this way is a complete and proper {\it subspace} of the space of square 
integrable functions over two variables. The space of functions 
$\{\psi(p,q)\equiv\<p,q|\psi\>\}$ defined for all $|\psi\>$ and all 
$(p,q)$ constitutes the {\it coherent state functional representation} 
of a one-particle Hilbert space. In this representation, for example, 
the basic Heisenberg operators are given by
  \bn P \ra -i\hbar\frac{\d}{\d q}\;, \hskip2cm Q\ra q+i\hbar
\frac{\d}{\d p}\;, \en
and thus the coherent state representation of the Schr\"odinger equation 
is given by
  \bn i\hbar\s\s\d\psi(p,q,t)/\d t=\H(\s-i\hbar\s\d/\d q,q+i
\hbar\s\d/\d p\s)\s\psi(p,q,t)\;. \en
What makes this a one-particle problem (rather than a special kind of 
two-particle problem) is the proper choice of acceptable initial conditions 
for this equation. In particular, it suffices to take as an initial 
condition the coherent state overlap function $\<p,q|p',q'\>$, for 
arbitrary $p'$ and $q'$. 

The solution to this form of Schr\"odinger's equation can be expressed 
in the form
  \bn \psi(p'',q'',t'')=\int K(p'',q'',t'';p',q',t')\s\psi(p',q',t')
\,dp'\s dq'/(2\pi\hbar)\;, \en
where $K(p'',q'',t'';p',q',t')$ denotes the propagator in the coherent 
state representation.

The coherent state representation differs from the usual Schr\"odinger 
representation given in (1), and it also has a different physical 
interpretation as well. For one thing, the meaning of $|\psi(x)|^2$ 
is the probability density to find the particle at position $x$. The 
meaning of $|\psi(p,q)|^2$, however, is rather different; this quantity 
is simply the probability that the state $|\psi\>$ can be found in the 
state $|p,q\>$. The physical meaning of the variable $x$ in the 
Schr\"odinger representation is that of a sharp position. Correspondingly, 
we may ask what is the physical meaning of the parameters $p$ and $q$? 
To answer that question it is convenient to restrict the choice of the 
fiducial vector very slightly by imposing the conditions that
$\<\eta|P|\eta\>=\<\eta|Q|\eta\>=0$. In that case, it follows that
  \bn \<p,q|P|p,q\>=p\;, \hskip2cm \<p,q|Q|p,q\>=q\;, \en
implying that unlike $x$ in the Schr\"odinger representation, the variables
$p$ and $q$ represent {\it mean values} in the coherent states. It is 
for this reason that we can specify both values simultaneously at the 
same time, something that could not be done if instead they both had 
represented sharp eigenvalues.

The propagator for the coherent state representation of Schr\"odinger's 
equation can also be given a formal phase space path integral form, namely
  \bn K(p'',q'',t'';p',q',t')={\cal M}\int \exp\{(i/\hbar)
\tint[\s p\s{\dot q}-H(p,q)]\s dt\}\;{\cal D}p\,{\cal D}q\;, \en
which superficially is just the same expression as (32)! What makes these 
expressions different is what values are pinned and more explicitly what 
are the respective lattice space formulations. In the coherent state case, 
we have
  \bn&& \hskip-.5cm K(p'',q'',t'';p',q',t')=\lim_{\ep\ra0}
\int\cdots\int\exp\{(i/\hbar)\s\Sigma_{l=0}^N
[\s\half\s(p_{l+1}+p_l)(q_{l+1}-q_l)\no\\
&&\hskip.5cm-\ep\s H(\half(p_{l+1}+p_l)+i\half(q_{l+1}-q_l),
\half(q_{l+1}+q_l)-i\half(p_{l+1}-p_l))\s]\s\}\no\\
&&\hskip.2cm\times\exp\{-(1/4\hbar)\Sigma_{l=0}^N[(p_{l+1}-p_l)^2+
(q_{l+1}-q_l)^2\s]\}\;\Pi_{l=1}^N\,dp_l\s dq_l/(2\pi\hbar)\;. \en
Observe that there are the same number of $p$ and $q$ integrations in 
this expression. Such a conclusion is fully in accord with the combination 
law as expressed in the coherent state representation, namely
\bn &&\hskip-.6cm K(p''',q''',t''';p',q',t')\no\\
&&\hskip.2cm=\int
K(p''',q''',t''';p'',q'',t'')\,K(p'',q'',t'';p',q',t')\,dp''\s
dq''/(2\pi\hbar)\;.  \en 

\section*{Daubechies and Klauder, 1985}
By any measure, phase space is the natural arena for both classical and 
quantum mechanics. In classical physics, dynamics evolves in phase space 
in a highly symmetric fashion for general Hamiltonians, while in quantum 
physics both $P$ and $Q$ appear in remarkably similar roles within the 
general formalism. Moreover, as stressed previously, the phase space 
path integral is more widely applicable than the original configuration 
space path integral.

In 1985, Daubechies and Klauder \cite{dau} examined the phase space 
path integral in a new light. In particular, they were led to consider 
the expression
  \bn \lim_{\nu\ra\infty}{\cal M}_\nu\int \exp\{{\textstyle\frac{i}{\hbar}}
\tint[p\s{\dot q}-H(p,q)]\s dt\}\,\exp\{-{\textstyle\frac{1}{2\nu}}
\tint[{\dot p}^2+{\dot q}^2]\s dt\}\,{\cal D}p\,{\cal D}q\;,\en
which extends the continuous-time regularization procedure to phase 
space path integrals. Let us initially look at this expression from a 
general viewpoint. Evidently we can interpret the regularization as two 
independent Brownian motion regularizations in the limit as the diffusion 
constant diverges, which is rather like the original Gel'fand-Yaglom 
procedure. Alternatively, and since from a classical point of view $p$ 
is often rather like ${\dot q}$ (depending on $H$, of course), this 
procedure has some features superficially in common with that of It\^o. 
However, there are also important differences with those previous 
procedures as well. The chosen regularization ensures that both $p$ and 
$q$ are continuous functions but that ${\dot p}$ and ${\dot q}$ are 
almost nowhere defined. The derivatives enter the integrand only in the 
form $\tint p\s{\dot q}\s dt$. However, this term can be interpreted as 
$\tint p\s dq$, which for Brownian motion paths is a standard stochastic 
integral that has been  well studied in both its It\^o and Stratonovich 
versions; for reasons to be given below, when it becomes important to 
choose a rule we shall choose the Stratonovich rule. We interpret the 
integration in (56) to be pinned for both $p$ and $q$ at both end points: 
specifically, $(p(t''),q(t''))=(p'',q'')$ and $(p(t'),q(t'))=(p',q')$. In 
this case it is clear that the result of (56) yields a function of the form
  \bn  K(p'',q'',t'';p',q',t')\;. \en

By the time of 1985, representations of propagators for one-particle 
quantum systems in the form of coherent state representations, such as 
those described in an earlier section, were well known to many researchers. 
If, indeed, (56) turned out to yield coherent state representations for 
solutions to the Schr\"odinger equation --- as opposed, say, to either the 
$q$-space or $p$-space representations --- there would be no reason for 
concern. In fact, that is exactly what happens!

In \cite{dau} it was rigorously established that 
\bn &&\lim_{\nu\ra\infty}{\cal M}_\nu\int e^{(i/\hbar)
\tint[p{\dot q}-H(p,q)]\s dt}\s e^{-(1/2\nu)
\tint[{\dot p}^2+{\dot q}^2]\s dt}\,{\cal D}p\,{\cal D}q\no\\
&&\hskip1cm=\lim_{\nu\ra\infty}2\pi\hbar\, e^{\nu T/2\hbar}\int
e^{(i/\hbar)\tint[p\s dq-H(p,q)\s dt]}\,d\mu_W^\nu(p,q)\no\\
&&\hskip1cm\equiv\<p'',q''|\s e^{-i(t''-t')\H/\hbar\s}\s|p',q'\>\no\\
&&\hskip1cm\equiv K(p'',q'',t'';p',q',t')\;,  \en
where the second line of (58) is a mathematically rigorous formulation of the
heuristic and formal first line, and $\mu_W^\nu$ denotes pinned Wiener
measure.
In fact, the connection indicated here is far reaching in that the choice of 
regularization also determines that 
 \bn && |p,q\>\equiv e^{-iqP/\hbar}\s e^{ipQ/\hbar}\s|0\>\;,\\
  && [Q,P]=i\hbar I\;,\hskip1cm (Q+iP)\s|0\>=0\;, \\
  &&\H=\int H(p,q)\s|p,q\>\<p,q|\,dp\,dq/(2\pi\hbar)\;. \en
A sufficient set of technical assumptions ensuring the validity of this 
representation is given by
 \bn &&a)\hskip.5cm \int H(p,q)^2\, e^{-\alpha(p^2+q^2)}\,dp\s dq<\infty,\; 
{\rm for\;all}\;\alpha>0\;, \\
&&b)\hskip.5cm \int H(p,q)^4 e^{-\beta(p^2+q^2)}\,dp\s dq<\infty,\;
{\rm for\;some}\;\beta<1/2\hbar\;,\\
  &&c)\hskip.5cm \H {\rm \;is \;essentially \;self \;adjoint \;on \;the \;
span \;of \;finitely \;many}\no\\
&&\hskip.7cm {\rm \;number \;eigenstates}\;.  \en\vskip.2cm
A wide class of Hamiltonians --- but by no means all acceptable such 
operators --- consists of all semibounded, symmetric (Hermitian) 
polynomials of $P$ and $Q$. We also note that the connection between 
the operator $\H$ and its symbol $H(p,q)$ given in (61) is that of 
anti-normal ordering. In particular, this means that 
 \bn e^{-(r-is)(Q+iP)/2\hbar}\s e^{(r+is)(Q-iP)/2\hbar}=
\int e^{i(sq-rp)/\hbar}|p,q\>\<p,q|\,dp\s dq/(2\pi\hbar)\;, \en
and any other connection follows by suitable linear combinations. As 
an alternative association we also note that
  \bn H(p,q)= e^{-(\hbar/4)(\d_p^2+\d_q^2)}\,H_W(p,q)\;,\en
where $H_W(p,q)$ denotes the well-known Weyl symbol of the operator 
$\H$. [In
most of the author's earlier work, the particular symbol $H(p,q)$ used 
here has been denoted by the lower case symbol $h(p,q)$.]

Observe that the auxiliary factor in (56) that provides  the regularization 
involves a {\it metric}, specifically, $d\sigma^2=d\sigma(p,q)^2=dp^2+dq^2$, 
on classical phase space. Such a metric characterizes a flat, 
two-dimensional phase space expressed in Cartesian coordinates.

It is interesting to consider the transformation of (58) under a canonical 
change of phase space coordinates. To this end we introduce 
${\o p}={\o p}(p,q)$ and
${\o q}={\o q}(p,q)$ as two new classical canonical coordinates that are 
related to the original canonical coordinates by the relation
  \bn {\o p}\s d{\o q}=p\s dq+dF({\o q},q)\;, \en
where $F$ is known as the generator of the transformation. This is a standard
form for this kind of relation in which ${\o q}$ and $q$ are regarded as 
the independent variables; another version that we will find more useful 
is given by
  \bn {\o p}\s d{\o q}+d{\o G}({\o p},{\o q})=p\s dq\;. \en
Under such a canonical coordinate transformation, the metric 
$d\sigma^2=dp^2 +dq^2$ assumes the form 
 \bn d\sigma^2=d\sigma({\o p},{\o q})^2=A({\o p},{\o q})\s 
d{\o p}^2+2B({\o p},{\o q})\s d{\o p}\s d{\o q}+C({\o p},{\o q})
\s d{\o q}^2\;,  \en
for suitable $A$, $B$, and $C$, but, nevertheless, the metric still 
characterizes the same flat, two-dimensional phase space even though it 
is now expressed, generally speaking, in curvilinear canonical coordinates.

In the original Cartesian coordinates the stochastic integral $\tint p\s dq$
evaluated by the Stratonovich rule is identical to its evaluation using the
It\^o rule. Under coordinate transformations, we adopt the Stratonovich 
definition [mid-point rule, cf.(54)] since for this rule the differential 
and integral laws of ordinary calculus still apply. Consequently, the 
differential rules (67) and (68) which apply for classical functions 
(say $C^2$), apply equally well to Brownian paths (which are only $C^0$). 
Recognizing that the Hamiltonian function transforms as a scalar, we see that
  \bn {\o H}({\o p},{\o q})\equiv H(p({\o p},{\o q}),q({\o p},{\o q}))
=H(p,q)\;.  \en

Assembling the separate parts, we learn that under a canonical coordinate 
transformation, (58) becomes
  \bn &&\lim_{\nu\ra\infty}{\cal M}_\nu\int e^{(i/\hbar)\tint
[{\o p}\s{\dot{\o q}}+{\dot{\o G}}({\o p},{\o q})-{\o H}({\o p},{\o q})]
\s dt}\,
e^{-(1/2\nu)\tint[d\sigma({\o p},{\o q})^2/dt^2]\,dt\s}\;{\cal D}{\o p}\,
{\cal D}{\o q}\no\\
&&\hskip1cm=\lim_{\nu\ra\infty} 2\pi\hbar \,e^{\nu T/2\hbar}\int 
e^{(i/\hbar)\tint[{\o p}\s d{\o q}+d{\o G}({\o p},{\o q})-{\o H}({\o p},
{\o q})\s dt]}\;d{\o\mu}_W^\nu({\o p},{\o q})\no\\
&&\hskip1cm\equiv\<{\o p}'',{\o q}''|\s e^{-i(t''-t')\H/\hbar\s}\s|{\o p}',
{\o q}'\>\no\\
&&\hskip1cm\equiv {\o K}({\o p}'',{\o q}'',t'';{\o p}',{\o q}',t') \;, \en
where ${\o\mu}_W^\nu$ denotes pinned Wiener measure on a two-dimensional flat
phase space expressed in curvilinear coordinates. The term
  \bn \tint d{\o G}({\o p},{\o q})={\o G}({\o p}'',{\o q}'')-{\o G}({\o p}',
{\o q}')\;,  \en
and thus its presence amounts to no more than a phase change of the 
corresponding coherent states. Specifically, 
 \bn && |{\o p},{\o q}\>\equiv e^{-i{\o G}({\o p},{\o q})/\hbar}\s 
e^{-ip({\o p},{\o q})Q/\hbar}\s e^{iq({\o p},{\o q})P/\hbar}\s|0\>=
e^{-i{\o G}({\o p},{\o q})/\hbar}\s|p,q\>\;,\\
 &&[Q,P]=i\hbar I\;,\hskip1cm (Q+iP)\s|0\>=0\;,\\
&&\H=\int {\o H}({\o p},{\o q})\s|{\o p},{\o q}\>\<{\o p},{\o q}|
\,d{\o p}\s d{\o q}/(2\pi\hbar)\;. \en
Under a canonical coordinate transformation, therefore, and apart from a 
possible phase change, observe that the coherent state vectors are 
{\it unchanged}; it is only their {\it labels} that have changed. Observe 
further that the Hamiltonian operator $\H$ is completely unchanged; only 
the details of its representation are different in the new coordinate system.

In summary, we see that a continuous-time, Brownian motion regularization of 
the phase space path integral can be rigorously established. It applies to a 
wide
class of Hamiltonians, and the formulation is fully covariant under general 
canonical coordinate transformations. The only additional ingredient to 
attain this covariance is the introduction of a {\it metric}, $d\sigma^2$, 
on classical phase space which is ultimately used to support the Wiener 
measure regularization. As a general rule, it is noteworthy that 
quantization schemes that agree with orthodox quantum mechanics --- and 
therefore agree with a vast number of experiments --- do {\it not} exhibit 
manifest covariance under canonical coordinate transformations. Hence, the 
fact that the present procedure exhibits covariance under canonical 
coordinate transformations is especially to be welcomed.

As a final comment, we remind the reader of the general requirement that 
canonical quantization, i.e., the choice of certain phase space variables 
to be ``promoted'' to operators, should be carried out in Cartesian 
coordinates \cite{dir}. This rule has caused much concern over the years 
and its elimination has been, partially at least, a motivating factor in 
certain alternative quantization schemes. Instead of seeking to eliminate 
the Cartesian coordinates, we fully accept them, and, indeed, we 
understand the need for them as follows \cite{und}: With only a symplectic 
form on phase space --- and in the absence of any phase space 
metric --- there is no self-referential mechanism to associate the 
correct {\it physical meaning} of any given {\it mathematical expression}, 
say, for the Hamiltonian $H(p,q)$. In particular, if $H(p,q)=p^2/2$, and 
in the absence of any further information, how is one to determine that this 
particular expression is the Hamiltonian for a free particle, or for an 
harmonic oscillator, or for some particular anharmonic oscillator, etc., 
any one of which can be brought to that mathematical form by a suitable 
canonical coordinate transformation. It is with the addition of a 
flat-space phase-space metric, and with due regard to its coordinatized 
expression, that the physical significance of a given mathematical 
expression can be determined, and with that determination, the
quantization procedure can lead to the correct energy spectrum for the 
physical system under consideration. The phase space continuous-time 
regularization scheme discussed in this section offers the great 
advantage that the introduction of the metric into the formal phase 
space path integral simultaneously gives mathematical and physical 
meaning to the path integral by assuring that the proper interpretation of 
the Hamiltonian is maintained throughout the quantization procedure. In 
point of fact, one may say that the procedure we have illustrated offers 
a true {\it geometric quantization procedure}, and consequently, the entire 
procedure can therefore be expressed in a {\it coordinate free form}. That 
is indeed the case, and the interested reader is referred to \cite{kla3} 
for further discussion of this point.

\subsection*{Final remarks}
A number of further developments and applications regarding 
continuous-time regularizations have appeared in recent years. We do not 
enter into any details of these works here, but rather list just a few 
additional papers for further study: \cite{pau,mar,aff1}.

\section*{Dedication}
It is a distinct pleasure to dedicate this paper to my long time friend 
and colleague, Hiroshi Ezawa. I have known Hiroshi for approximately 
forty years and I have profited greatly from the many times we have been 
together and during which we have often worked closely on a variety of 
topics of mutual interest. I wish him many more years of good health and 
productive research. 

\section*{Acknowledgments}
Thanks are expressed to Lorenz Hartmann and Dae-Yup Song for their comments
on the manuscript.


\begin{thebibliography}{99}

\bibitem{fey} R.P.~Feynman, Space-time Approach to Nonrelativistic Quantum
Mechanics, {\it Rev.~Modern Phys.} {\bf 20}, 367-387 (1948).

\bibitem{kac}M.~Kac, On Some Connection between Probability Theory and 
Differential and Integral Equations, {\it Proc.~2nd Berkeley 
Sympos.~Math.~Stat. and Prob.}, 189-215 (1951). Reprinted in {\it Mark Kac: 
Probability, Number Theory, and Statistical Physics  -- Selected Papers}, 
Eds. K.~Baclawski and M.D.~Donsker, (The MIT Press, Cambridge, MA, 1979).

\bibitem{gel}I.M.~Gel'fand and A.M.~Yaglom,  Integration in Function Spaces 
and its Applications in Quantum Physics, {\it Uspekhi Mat.~Nauk} vol. II, 
77-114 (1956) (in russian); {\it J.~Math.~Phys.} {\bf 1}, 48-69 (1960) 
(English translation).  

\bibitem{cam}R.H.~Cameron, A Family of Integrals Serving to Connect the 
Wiener and Feynman Integrals, {\it Journal of Math.~and Phys.} {\bf 39}, 
126-140 (1960); The Ilstow and Feynman Integrals, {\it J.~Analyse Math.} 
{\bf 10}, 287-361 (1962/63).

\bibitem{ito}K.~It\^o, Wiener Integral and Feynman Integral, {\it Proc. 
Fourth Berkeley Symp. on Math., Stat. and Prob.} {\bf 2}, 227-238 (1960); 
Generalized Uniform Complex Measures in the Hilbertian Metric Space with 
their Application to the Feynman Integral, {\it Proc. Fifth Berkeley Symp. 
on Math., Stat. and Prob.} {\bf 2}, 145-161 (1965). Both articles are 
reprinted in {\it Kiyosi It\^o, Selected Papers}, Eds. D.W.~Stroock 
and S.R.S.~Varadhan, (Springer-Verlag, New York, 1987).

\bibitem{orn}See, e.g., G.~Roepstorff, {\it Path Integral Approach to 
Quantum Physics}, (Springer-Verlag, Berlin, 1996).

\bibitem{fey2}R.P.~Feynman, An Operator Calculus having Applications in 
Quantum Electrodynamics, {\it Phys. Rev.} {\bf 84}, 108-128 (1951).

\bibitem{klacoh}See, e.g., J.R.~Klauder, The Action Option and the Feynman 
Quantization of Spinor Fields in Terms of Ordinary $c$-Numbers, 
{\it Ann.~Phys.} {\bf 11}, 123-168 (1960); V.~Bargmann, On a Hilbert 
Space of Analytic Functions and an Associated Integral Transform. Part. I. 
{\it Commun.~Pure Appl.~Math.} {\bf 14}, 187-214 (1961); I.E.~Segal, 
Mathematical Characterization of the Physical Vacuum for a Linear 
Bose-Einstein Field, {\it Illinois J.~Math.} {\bf 6}, 500-523 (1962); 
J.R.~Klauder, Continuous-Representation Theory I.  Postulates of Continuous 
Representation Theory, {\it J.~Math.~Phys.} {\bf 4}, 1055-1058 (1963); 
J.~McKenna and J.R.~Klauder, Continuous-Representation Theory IV.  
Structure of a 
Class of Function Spaces Arising from Quantum Mechanics, 
{\it J.~Math.~Phys.} 
{\bf 5}, 878-896 (1964); {\it Coherent States:  Applications to Physics 
and 
Mathematical Physics}, Eds. J.R.~Klauder and B.-S.~Skagerstam 
(World Scientific, Singapore, 1985).

\bibitem{dau}I.~Daubechies and J.R.~Klauder, Quantum Mechanical Path 
Integrals with Wiener Measures for All Polynomial Hamiltonians. II, 
{\it J.~Math.~Phys.} {\bf 26}, 2239-2256 (1985).
 
\bibitem{dir}P.A.M.~Dirac, {\it The Principles of Quantum Mechanics} 
(Oxford
University Press, Oxford, Third Edition, 1947), p.114.

\bibitem{und}J.R.~Klauder, Understanding Quantization, {\it Found.~Phys.} 
{\bf 27}, 1467-1483 (1997).

\bibitem{kla3}J.R.~Klauder, Quantization {\it Is} Geometry, After All,
{\it Ann.~Phys.} {\bf 188}, 120-141 (1988). 
 
\bibitem{pau}I.~Daubechies, J.R.~Klauder, and T.~Paul, Wiener Measures for 
Path Integrals with Affine Kinematic Variables, {\it J.~Math.~Phys.} 
{\bf 28}, 85-102 (1987). 

\bibitem{mar}P.~Maraner, Landau Ground State on Riemannian Surfaces, 
{\it Mod. Phys.~Lett.} A {\bf 7}, 2555-2558 (1992).

\bibitem{aff1}J.R.~Klauder, Noncanonical Quantization of Gravity. I. 
Foundations of Affine Quantum Gravity, {\it J.~Math.~Phys.} {\bf 40}, 
5860-5882 (1999); Noncanonical Quantization of Gravity. II. Constraints 
and the Physical Hilbert Space, {\it J.~Math.~Phys.} {\bf 42}, 4440-4464 
(2001);
The Affine Quantum Gravity Programme, {\it  Class.~Quant.~Grav.} 
{\bf 19}, 817-826 (2002).
\end{thebibliography}
\end{document}